\documentclass{sig-alternate}
\usepackage{amsfonts}
\usepackage{listings}
\usepackage{amsmath}
\usepackage{amssymb}
\usepackage{graphicx}
\usepackage{algorithm}
\usepackage{url}
\usepackage[noend]{algpseudocode}
\newcommand{\ignore}[1]{}

\begin{document}

\title{Monoidify!\ Monoids as a Design Principle for\\Efficient MapReduce Algorithms}

\author{\alignauthor Jimmy Lin \\
\affaddr{University of Maryland}\\
\affaddr{jimmylin@umd.edu}}

\date{}

\maketitle

\numberofauthors{1}
\additionalauthors{}

\section{Introduction}

The purpose of this short paper is to share a recent observation I
made in the context of my introductory graduate course on MapReduce at
the University of Maryland. It is well known that since the
sort/shuffle stage in MapReduce is costly, local aggregation is one
important principle to designing efficient algorithms. This typically
involves using combiners or the so-called in-mapper combiner
technique~\cite{Lin_Dyer_book}. However, can we be more precise in
formulating this design principle for pedagogical purposes? Simply
saying ``use combiners'' or ``use in-mapper combining'' is
unsatisfying because it leaves open the obvious question of {\it how}?
What follows is my attempt to formulate a more precise design
principle in terms of monoids---the idea is quite simple, but I
haven't seen anyone else make this observation before in the context
of MapReduce.

Let me illustrate with a running example I often use to illustrate
MapReduce algorithm design, which is detailed in Lin and
Dyer~\cite{Lin_Dyer_book}. Given a large number of key--value pairs
where the keys are strings and the values are integers,
we wish to find the average of all the values by key. In SQL, this is
accomplished with a simple group-by and \textsc{Avg}. Here is the
na\"ive MapReduce algorithm:

\begin{algorithm}[h]
\caption{}
\label{algorithm:chapter3:average}
\algrenewcommand\algorithmicfunction{\textbf{class}}
\algrenewcommand\algorithmicprocedure{\textbf{method}}
  \begin{algorithmic}[1]
    \Function{Mapper}{}
    \Procedure{Map}{$\textrm{string }t, \textrm{integer }r$}
    \State $\textsc{Emit}(t, r)$
    \EndProcedure
    \EndFunction
  \end{algorithmic}

  \begin{algorithmic}[1]
    \Function{Reducer}{}
    \Procedure{Reduce}{$\textrm{string }t, \textrm{integers }[ r_1, r_2, \ldots ]$}
    \State $sum \gets 0$
    \State $cnt \gets 0$
    \ForAll{$ r \in [ r_1, r_2, \ldots ]$}
    \State $sum \gets sum + r$
    \State $cnt \gets cnt + 1$
    \EndFor
    \State $r_{avg} \gets sum/cnt$
    \State $\textsc{Emit}(t, r_{avg})$
    \EndProcedure
    \EndFunction
  \end{algorithmic}
\end{algorithm}

This isn't a particularly efficient algorithm because the mappers do no
work and all data are shuffled (across the network) over to the reducers.
Furthermore, the reducer cannot be used as a combiner.
Consider what would happen if we did:\ the combiner would compute the
mean of an arbitrary subset of values with the same key,
and the reducer would compute the mean of those values. As a concrete
example, we know that:
\begin{displaymath}
\textsc{Avg}(1, 2, 3, 4, 5) \ne \textsc{Avg}( \textsc{Avg}(1, 2), \textsc{Avg}(3, 4, 5))
\end{displaymath}

\noindent In general, the mean of means of arbitrary subsets of a set
of values is not the same as the mean of the set of values.

So how might we properly take advantage of combiners?  An attempt is
shown in Algorithm~\ref{algorithm:chapter3:average-fail}.

\begin{algorithm}[h]
\caption{}
\label{algorithm:chapter3:average-fail}
\algrenewcommand\algorithmicfunction{\textbf{class}}
\algrenewcommand\algorithmicprocedure{\textbf{method}}
  \begin{algorithmic}[1]
    \Function{Mapper}{}
    \Procedure{Map}{$\textrm{string }t, \textrm{integer }r$}
    \State $\textsc{Emit}(\textrm{string }t, \textrm{integer }r)$
    \EndProcedure
    \EndFunction
  \end{algorithmic}

  \begin{algorithmic}[1]
    \Function{Combiner}{}
    \Procedure{Combine}{$\textrm{string }t, \textrm{integers }[ r_1, r_2, \ldots ]$}
    \State $sum \gets 0$
    \State $cnt \gets 0$
    \ForAll{$ \textrm{integer }r \in \textrm{integers }[ r_1, r_2, \ldots ]$}
    \State $sum \gets sum + r$
    \State $cnt \gets cnt + 1$
    \EndFor
    \State $\textsc{Emit}(\textrm{string }t, \textrm{pair } (sum, cnt))$
    \EndProcedure
    \EndFunction
  \end{algorithmic}

  \begin{algorithmic}[1]
    \Function{Reducer}{}
    \Procedure{Reduce}{$\textrm{string }t, \textrm{pairs }[ (s_1, c_1), (s_2, c_2) \ldots ]$}
    \State $sum \gets 0$
    \State $cnt \gets 0$
    \ForAll{$ \textrm{pair }(s, c) \in \textrm{pairs }[ (s_1, c_1), (s_2, c_2) \ldots ]$}
    \State $sum \gets sum + s$
    \State $cnt \gets cnt + c$
    \EndFor
    \State $r_{avg} \gets sum/cnt$
    \State $\textsc{Emit}(\textrm{string }t, \textrm{integer } r_{avg})$
    \EndProcedure
    \EndFunction
  \end{algorithmic}
\end{algorithm}

The mapper remains the same, but we have added a combiner that
partially aggregates results by separately tracking the numeric components
necessary to arrive at the mean.  The combiner receives each string
and the associated list of integers, from which it computes the
sum of those values and the number of integers encountered (i.e., the count).
The sum and count are packaged into a pair and emitted as
the output of the combiner, with the same string as the key.  In the
reducer, pairs of partial sums and counts can be aggregated to arrive
at the mean.

The problem with this algorithm is that it doesn't actually work.
Combiners must have the same input and output key--value type, which
also must be the same as the mapper output type and the reducer input
type. This is clearly not the case. To understand why this restriction
is necessary, remember that combiners are
optimizations that cannot change the correctness of the algorithm.  So
let us remove the combiner and see what happens:\ the output value
type of the mapper is integer, so the reducer should receive a
list of integers. But the reducer actually expects a list of
pairs! The correctness of the algorithm is contingent on the combiner
running on the output of the mappers, and more specifically, that the
combiner is run exactly once.  Hadoop, for example, makes no
guarantees on how many times combiners are called; it could be zero,
one, or multiple times. This algorithm violates the MapReduce
programming model.

Another stab at the solution is shown in
Algorithm~\ref{algorithm:chapter3:average-efficient}:

\begin{algorithm}[h]
\caption{}
\label{algorithm:chapter3:average-efficient}
\algrenewcommand\algorithmicfunction{\textbf{class}}
\algrenewcommand\algorithmicprocedure{\textbf{method}}
  \begin{algorithmic}[1]
    \Function{Mapper}{}
    \Procedure{Map}{$\textrm{string }t, \textrm{integer }r$}
    \State $\textsc{Emit}(t, (r, 1))$
    \EndProcedure
    \EndFunction
  \end{algorithmic}

  \begin{algorithmic}[1]
    \Function{Combiner}{}
    \Procedure{Combine}{$\textrm{string }t, \textrm{pairs }[ (s_1, c_1), (s_2, c_2) \ldots ]$}
    \State $sum \gets 0$
    \State $cnt \gets 0$
    \ForAll{$(s, c) \in [ (s_1, c_1), (s_2, c_2) \ldots ]$}
    \State $sum \gets sum + s$
    \State $cnt \gets cnt + c$
    \EndFor
    \State $\textsc{Emit}(t, (sum, cnt))$
    \EndProcedure
    \EndFunction
  \end{algorithmic}

  \begin{algorithmic}[1]
    \Function{Reducer}{}
    \Procedure{Reduce}{$\textrm{string }t, \textrm{pairs }[ (s_1, c_1), (s_2, c_2) \ldots ]$}
    \State $sum \gets 0$
    \State $cnt \gets 0$
    \ForAll{$(s, c) \in [ (s_1, c_1), (s_2, c_2) \ldots ]$}
    \State $sum \gets sum + s$
    \State $cnt \gets cnt + c$
    \EndFor
    \State $r_{avg} \gets sum/cnt$
    \State $\textsc{Emit}(t, r_{avg})$
    \EndProcedure
    \EndFunction
  \end{algorithmic}
\end{algorithm}

The algorithm is now correct. In the mapper we emit as the
intermediate value a pair consisting of the integer and one---this
corresponds to a partial count over one instance. The combiner
separately aggregates the partial sums and the partial counts (as
before), and emits pairs with updated sums and counts. The reducer is
similar to the combiner, except that the mean is computed at the end.
In essence, this algorithm transforms a non-associative operation
(mean of values) into an associative operation (element-wise sum of a
pair of numbers, with a division at the end).

Finally, Algorithm~\ref{algorithm:chapter3:average-more-efficient}
shows an even more efficient algorithm that exploits the in-mapper
combining pattern:

\begin{algorithm}[h]
\caption{}
\label{algorithm:chapter3:average-more-efficient}
\algrenewcommand\algorithmicfunction{\textbf{class}}
\algrenewcommand\algorithmicprocedure{\textbf{method}}
  \begin{algorithmic}[1]
    \Function{Mapper}{}
    \Procedure{Initialize}{{}}
      \State $S \gets \textrm{new }\textsc{AssociativeArray}$
      \State $C \gets \textrm{new }\textsc{AssociativeArray}$
    \EndProcedure
    \Procedure{Map}{$\textrm{string }t, \textrm{integer }r$}
      \State $S\{t\} \gets S\{t\} + r$
      \State $C\{t\} \gets C\{t\} + 1$
    \EndProcedure
    \Procedure{Close}{{}}
    \ForAll{$\textrm{term }t \in S$}
      \State $\textsc{Emit}(\textrm{term }t, \textrm{pair }(S\{t\}, C\{t\}))$
    \EndFor
    \EndProcedure
    \EndFunction
  \end{algorithmic}
\end{algorithm}

Inside the mapper, the partial sums and counts
associated with each string are held in memory across input key--value
pairs.  Intermediate key--value pairs are emitted only after the entire
input split has been processed; similar to before, the value is a pair
consisting of the sum and count.  The reducer is exactly the same as
in Algorithm~\ref{algorithm:chapter3:average-efficient}. Moving partial
aggregation from the combiner into the mapper assumes that
the intermediate data structures will fit into memory, which may not
be a valid assumption. However, in cases where the assumption holds,
the in-mapper combining technique can be substantially faster than using
normal combiners, primarily due to the savings in not needing to
materialize intermediate key--values pairs.

\section{Monoidify!}

Okay, what have we done to make this particular algorithm work? The
answer is that we've created a monoid out of the intermediate value!

How so? First, a recap on monoids:\ a monoid is an algebraic structure
with a single associative binary operation\footnote{\small In many
  cases the operation is commutative as well, so we
  actually have a commutative monoid, although in this paper I won't
  focus on this distinction (i.e., in many places where I refer to a
  monoid, to be more precise it's actually a commutative monoid).} and
an identity element. As a simple example, the natural numbers form a
monoid under addition with the identity element 0. Applied to our
running example, it's now evident that the intermediate value in
Algorithm~\ref{algorithm:chapter3:average-efficient} forms a monoid:\ the
set of all tuples of non-negative integers with the identity element
$(0,0)$ and the element-wise sum operation, $(a, b) \oplus (c, d) =
(a+c, b+d)$.

Thus, one principle for designing efficient MapReduce algorithms can
be precisely articulated as follows:\ create a monoid out of the
intermediate value emitted by the mapper. Once we ``monoidify'' the
object, proper use of combiners and the in-mapper combining techniques
becomes straightforward.\footnote{\small In
  Algorithm~\ref{algorithm:chapter3:average-more-efficient}, the
  elements of the tuple have been pulled apart and stored in separate
  data structures, but that's a specific implementation choice not
  germane to the design principle.}  This principle also explains why
the reducer in Algorithm~\ref{algorithm:chapter3:average} cannot be
used as a combiner and why
Algorithm~\ref{algorithm:chapter3:average-fail} doesn't
work.\footnote{\small This exposition glosses over the fact that at the end
  of the computation, we break apart the pair to arrive at the mean,
  which destroys the monoid, but this is a one-time termination
  operation that can be treated as ``post-processing''.}

\section{Other Examples}

The ``monoidify'' principle readily explains another MapReduce
algorithm I often use for pedagogical purposes:\ the problem of
building word co-occurrence matrices from large natural language
corpora, a common task in corpus linguistics and statistical natural
language processing. Formally, the co-occurrence matrix of a corpus is
a square $n \times n$ matrix where $n$ is the number of unique words
in the corpus (i.e., the vocabulary size). A cell $m_{ij}$ contains
the number of times word $w_i$ co-occurs with word $w_j$ within a
specific context---a natural unit such as a sentence, paragraph, or a
document, or a certain window of $m$ words (where $m$ is an
application-dependent parameter). Note that the upper and lower
triangles of the matrix are identical since co-occurrence is a
symmetric relation, though in the general case relations between words
need not be symmetric. For example, a co-occurrence matrix $M$ where
$m_{ij}$ is the count of how many times word $i$ was immediately
succeeded by word $j$ (i.e., bigrams) would not be symmetric.
Beyond simple co-occurrence counts, the MapReduce algorithm for this
task extends readily to computing relative frequencies and forms the
basis of more sophisticated algorithms such as those for
expectation-maximization (where we're keeping track of pseudo-counts
rather than actual observed counts).

The so-called ``stripes'' algorithm~\cite{Lin_Dyer_book} for
accomplishing the co-occurrence computation is as follows:

\begin{algorithm}[h]
\caption{}
\label{algorithm:chapter3:coocur:stripes}

\algrenewcommand\algorithmicfunction{\textbf{class}}
\algrenewcommand\algorithmicprocedure{\textbf{method}}
  \begin{algorithmic}[1]
    \Function{Mapper}{}
    \Procedure{Map}{$\textrm{docid }a, \textrm{doc }d$}
    \ForAll{$\textrm{term }w \in \textrm{doc }d$}
    \State $H \gets \textrm{new }\textsc{AssociativeArray}$
    \ForAll{$\textrm{term }u \in \textsc{Neighbors}(w)$}
    \State $H\{u\} \gets H\{u\} + 1$
    \EndFor
    \State $\textsc{Emit}(\textrm{Term }w, \textrm{ Stripe }H)$
    \EndFor
    \EndProcedure
    \EndFunction
  \end{algorithmic}

  \begin{algorithmic}[1]
    \Function{Reducer}{}
    \Procedure{Reduce}{$\textrm{term }w, \textrm{ stripes }[H_1, H_2, H_3,\ldots ]$}
    \State $H_f \gets \textrm{new }\textsc{AssociativeArray}$
    \ForAll{$\textrm{stripe }H \in \textrm{stripes }[H_1, H_2, H_3, \ldots ]$}
    \State $\textsc{Sum}(H_f,H)$
    \EndFor
    \State $\textsc{Emit}(\textrm{term }w, \textrm{stripe }H_f)$
    \EndProcedure
    \EndFunction
  \end{algorithmic}
\end{algorithm}

In this case, the reducer can also be used as a combiner because
associative arrays form a monoid under the operation of element-wise
sum with the empty associative array as the identity element.

Here's another non-trivial example:\ Lin and
Kolck~\cite{Lin_Kolcz_SIGMOD2012} advocate the use of stochastic
gradient descent (SGD) for scaling out the training of
classifiers. Viewed from this perspective, SGD ``works'' because the model
parameter (i.e., a weight vector for linear models) comprise a monoid under
incremental training.

Other examples of interesting monoids that are useful for large-scale
data processing are found in Twitter's Algebird
package.\footnote{\small \url{github.com/twitter/algebird}} These
include Bloom filters~\cite{Bloom_1970}, count-min
sketches~\cite{countmin}, hyperloglog counters~\cite{hyperloglog}.

Finally, it is interesting to note that regular languages form a
monoid under intersection, union, subtraction, and
concatenation. Since finite-state techniques are widely used in
computational linguistics and natural language processing, this
observation might hold implications for scaling out text processing
applications.

\section{Optimizations and Beyond}

In the context of MapReduce, it may be possible to elevate
``monoidification'' from a design principle (that requires manual
effort by a developer) to an automatic optimization that can be
mechanistically applied. For example, in Hadoop, one can imagine
declaring Java objects as monoids (for example, via an
interface). When these objects are used as intermediate values in a
MapReduce algorithm, some optimization layer can {\it automatically}
create combiners (or apply in-mapper combining) as appropriate.

The observation that monoids represent a design principle for
efficient MapReduce algorithms extends more broadly to large-scale
data processing in general. One concrete example is Twitter's
Summingbird project, which takes advantage of associativity to
integrate real-time and batch processing. The same monoid
(from Algebird, mentioned above) can be used
to hold state in a low-latency online application (i.e., operating on
an infinite stream) as well as in a scale-out batch processing job
(e.g., on Hadoop).

\section{Conclusions}

None of the ideas in this paper are
completely novel:\ the property of associativity and commutativity in
enabling combiners to work properly was pointed out in the original
MapReduce paper~\cite{Dean_Ghemawat_OSDI2004}. Independently, there
has been a recent resurgence of interest in functional programming and
its theoretical underpinnings in category theory. However, I haven't
seen anyone draw the connection between MapReduce algorithm design and
monoids in the way that I have articulated here---and therein lies the
small contribution of this piece:\ identifying a phenomenon and giving
it a name.

However, it remains to be seen whether this observation is actually
useful. Perhaps I am gratuitously introducing monoids just because
category theory is ``hip'' and in vogue. In a way, a monoid is simply
a convenient shorthand for saying:\ associative operations give an
execution framework great flexibility in sequencing computations, thus
allow opportunities for much more efficient execution. Thus, another
way to phrase the takeaway lesson is:\ take advantage of associativity
(and commutativity) to the greatest possible extent. This rephrasing
conveys the gist without needing to invoke references to algebraic
structures.

Finally, there remains the question of whether this observation is
actually useful as a pedagogical tool for teaching students how to
think in MapReduce (which was the original motivation for this
paper). It is often the case that introducing additional layers of
abstraction actually confuses students more than it clarifies
(especially in light of the previous paragraph). This remains an
empirical question I hope to explore in future offerings of my
MapReduce course.

To conclude, the point of this paper can be summed up in a pithy
directive:\ Go forth and monoidify!

\section{Acknowledgments}

I'd like to thank Chris Dyer and Oscar Boykin for helpful discussions
that have shaped the ideas discussed in this piece. Additional thanks
to Bill Howe and Jeffrey Ullman for comments on earlier drafts.

\bibliographystyle{abbrv}

\end{document}